\documentclass[english]{lni}

\usepackage{times}
\usepackage{epsfig}
\usepackage{amsmath}
\usepackage{amssymb}
\usepackage{mathtools, nccmath}
\usepackage{multirow}
\usepackage[table,xcdraw]{xcolor}
\usepackage{graphicx}
\usepackage{subfig}
\usepackage{fancyhdr}
\usepackage{listings} %
\usepackage{changepage} %

\fancypagestyle{titlepage}{
\fancyhead[RO]{\small N. Damer,  M. Gomez-Barrero,  K. Raja,  C. Rathgeb,  A.  Sequeira,\linebreak  M. Todisco,  and A. Uhl (Eds.): BIOSIG 2023, \linebreak Lecture Notes in Informatics (LNI), Gesellschaft f\"ur Informatik, Bonn 2023} %
\fancyfoot{}}

\renewcommand{\headrulewidth}{0.4pt} %
\title{Generalizability and Application of the Skin Reflectance Estimate Based on Dichromatic Separation (SREDS)}

 \author{ Joseph Drahos, Richard Plesh, Keivan Bahmani, Mahesh Banavar, Stephanie Schuckers
 \footnote{Department of Electrical and Computer Engineering, 8 Clarkson Ave, Potsdam, NY,\\
\{drahosj, pleshro, bahmank, mbanavar, sschucke\} @clarkson.edu
 \\This material is based upon work supported by the Center for Identification Technology Research and the National Science Foundation (NSF) under Grant No.$1650503$.
 }}

\begin{document}
\maketitle

\renewcommand{\refname}{References}
\setcounter{footnote}{2} %
\thispagestyle{titlepage}
\pagestyle{fancy}
\fancyhead{} %
\fancyhead[RO]{\small Generalizability and Application of SREDS \hspace{25pt}  \hspace{0.05cm}}
\fancyhead[LE]{\hspace{0.05cm}\small  \hspace{25pt} J Drahos, R Plesh, K Bahmani, M Banavar, and S Schuckers}
\fancyfoot{} %
\renewcommand{\headrulewidth}{0.4pt} %

\begin{abstract}
Face recognition (FR) systems have become widely used and readily available in recent history. However, differential performance between certain demographics has been identified within popular FR models. Skin tone differences between demographics can be one of the factors contributing to the differential performance observed in face recognition models. Skin tone metrics provide an alternative to self-reported race labels when such labels are lacking or completely not available e.g. large-scale face recognition datasets. 
In this work, we provide a further analysis of the generalizability of the Skin Reflectance Estimate based on Dichromatic Separation (SREDS) against other skin tone metrics and provide a use case for substituting race labels for SREDS scores in a privacy-preserving learning solution. Our findings suggest that SREDS consistently creates a skin tone metric with lower variability within each subject and SREDS values can be utilized as an alternative to the self-reported race labels at minimal drop in performance. Finally, we provide a publicly available and open-source implementation of SREDS to help the research community. Available at https://github.com/JosephDrahos/SREDS

\end{abstract}


\section{Introduction}
Face recognition systems are increasingly used as a form of biometric authentication for many new and existing systems. Research on the differential performance between demographics is an important topic of study to mitigate bias and ensure fairness \cite{FDR,GARBE}. Modern facial recognition systems use deep learning pipelines to take an image of a person's face and create a unique template for that person. In such systems, the demographic information of a dataset is needed to assess or mitigate the differential performance of a particular face recognition algorithm. However, many of the large-scale datasets which have been aggregated from public images on the internet and used to train and benchmark face recognition networks lack self-reported race labels. Additionally, the large scale of such datasets makes it impractical and expensive to efficiently label demographics by human annotators. As a result, methods to automatically label a dataset can provide a valuable asset. 

Our research focuses on the intersection of privacy preservation, bias mitigation, and skin tone metrics. We present our analysis of the Skin Reflectance Estimate based on Dichromatic Separation (SREDS) skin tone metric from \cite{sreds}. SREDS is a continuous skin tone metric that can be used to automatically label skin tones on face datasets. Our goal is to evaluate SREDS' ability to label datasets compared to other skin tone metrics, assess the generalizability of SREDS on unseen data, and demonstrate an application of SREDS using a sensitive information removal approach when race labels are not available.

\section{Background}

\subsection{Skin Tone Metrics}
\label{SKINTONE}
Previous methods formulated for generating a metric for subject skin tones to more accurately describe skin color are listed as follows: Fitzpatrick Skin Type (FST), Monk Skin Tone (MST) Scale, Individual Typology Angle (ITA), and Relative Skin Reflectance (RSR) \cite{FST, Monk_2019, ITA, RSR}. FST and its successor MST require a manual calculation from a survey of the subject, while ITA and RSR can be computed automatically via an algorithm. RSR was created to analyze skin tone for a specific dataset by fitting a Principal Component Analysis (PCA) model to the RGB space of the dataset. RSR is not resistant to changes in lighting and is specific to a particular dataset. The need for a skin tone metric that can be computed automatically and is more resistant to changes in lighting prompted the research that led to the Skin Reflectance Estimate based on a Dichromatic Separation (SREDS) \cite{sreds}. SREDS aims to decompose patches of skin into specular and diffuse components using the dichromatic reflectance model. A Kernal Principal Component Analysis (KPCA) is fit onto the diffuse components extracted from the dataset, resulting in a data-driven skin tone metric. 

\subsection{Bias Mitigation}
The inclusion of demographic information in a dataset is to observe and attempt to eliminate the differential performance between demographic groups in FR models. Differing methods of bias mitigation have been attempted and documented at the feature, comparison, and post-comparison levels. 
A method of bias mitigation at the feature level is the triplet mining approach of \cite{SERNA2022103682} which used a triplet loss for discrimination-aware learning. Closely related triplets are mined based on race information to try and train a new representation that mitigates biased learning within the face embedding space of a pre-trained model. 
At the comparison level, a learning classifier method reduces ethnic bias by introducing group and individual fairness to the decision process at the cost of matching performance \cite{terhorst2020comparison}. 
At the post-comparison level, an unsupervised method of score normalization has been presented to reduce bias between ethnic groups while increasing the performance of the system \cite{terhorst2020postcomparison}.

\subsection{Soft Biometric Privacy Preservation}
Soft biometric information such as gender, race, age, etc. is stored within the templates created from FR systems and can be extracted without the user's consent \cite{terhorst2020identity}. 
Methods of privacy preservation have been studied and introduced to protect users' sensitive information. The efforts in\cite{OthmanRoss_2015} produced a technique that morphed the input face with another face to mask the soft biometrics while maintaining matching performance. Another technique that added a perturbing element to the initial face image that would mask sensitive information while maintaining performance is \cite{MirjaliliRoss_2017}. 
Information removal networks attempt to remove sensitive information from the feature embedding space of the FR deep network. These methods require complex loss functions to maintain the performance accuracy of the network while also suppressing the racial information from the learned space, as performed in \cite{xu2018fairgan}. A method that combines the methods from \cite{SERNA2022103682, xu2018fairgan} and was used within this research is \cite{morales2020sensitivenets}, which attempts to maintain the inter-identity distance using triplet loss and simultaneously unlearn\footnote[2]{The term unlearn will be used throughout the paper in the same context as introduced in the literature \cite{morales2020sensitivenets}.} the facial features used to differentiate between demographic classes.

\begin{figure}[ht]
    \centering
    \subfloat[Intra-subject variation between diffuse and specular components extracted using SREDS.]{\includegraphics[width=7cm]{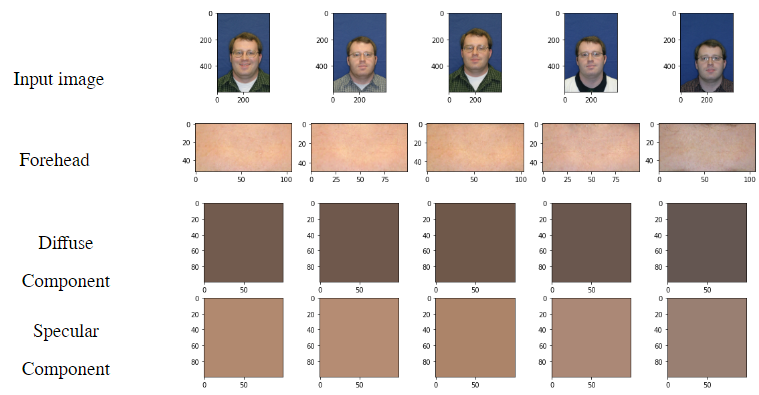}\label{Intrasubject Variation}}
    \subfloat[SensitiveNets model implementation]{\includegraphics[width=5cm]{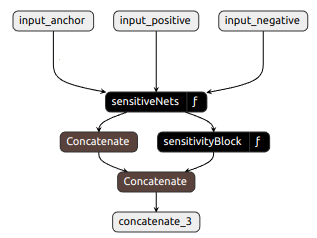}\label{snetsModelImpl}}
    \caption{}
    \label{fairness plots}
\end{figure}

\section{Methodology}

\subsection{Skin Tone Metrics Assessment}

The skin tone metrics outlined in section \ref{SKINTONE} will be used as a baseline to compare the previously developed methods to the performance of the SREDS measure. Individual typology angle (ITA) is a type of colorimetric analysis designed to measure acquired tanning \cite{ITA}. An RGB image is converted into CIE-Lab space \cite{RGBtoCIELAB}, as follows: (1) the ‘$L$’ component which quantifies luminance, (2) the ‘$a$’ component - absence or presence of redness, and (3) the ‘$b$’ component - yellowness. Using the ‘$L$’ and ‘$b$’ components, Pixel-wise ITA value, in degrees, can be estimated throughout an image as:

\begin{equation}
ITA = \frac{{\rm arctan}(L-50)}{b} * \frac{180}{\pi}. 
\end{equation}

To find suitable skin pixels in the image, a landmark extractor based on Dlib is used to detect the forehead, left cheek, and right cheek facial regions \cite{dlib}. For each facial region, ITA is computed over each pixel and smoothed using an averaging filter. The mode from each region’s resulting values is averaged to result in a single skin tone estimate for a face. 

Relative Skin Reflectance (RSR) is a process designed to relate the physical properties of the skin to the performance of facial recognition \cite{RSR}. The pipeline works by removing the confounding effects of imaging artifacts on skin pixels and fitting a line in the direction of the greatest variance in the RGB color space using PCA. The resulting metric is related to the skin tone of each subject relative to the rest of the photos in the dataset. Assumptions include consistent lighting, the same acquisition camera, and constant background. As a further limitation, the metric only indicates where a subject lies regarding net skin reflectance relative to the other subjects in the dataset, rather than an absolute measure.

The process to compute SREDS begins by extracting patches of skin from the forehead, right, and left cheeks using Dlib landmarks of each face image. Using the dichromatic reflection model as a guide, Non-Negative Matrix Factorization (NNMF) is used to estimate the diffuse and specular components of the selected skin patches. KPCA is utilized on the extracted diffuse components to learn a skin tone gradient across the dataset. The averaged value of the first principal components of the extracted diffuse bases for a particular face defines that person's SREDS score. The KPCA model used for SREDS is data-driven, so the generalizability of the KPCA model onto unseen datasets is a point of interest within this study. A full description of the extraction of SREDS is found in \cite{sreds}. 

\subsection{Datasets}
For our experiments, we selected datasets that included demographic information of subjects across race, age, gender, orientation, and lighting. We utilized  CMU Multi-PIE, MEDS-II, and Morph-II datasets \cite{multipie,meds2,morph}. Multi-PIE contains 750,000 sample images from 337 subjects images under 15 viewpoints with 19 illumination conditions. We selected three viewpoints (14 0, 05 1, 05 0) where full views of the face were captured for our testing, which reduced our sample images to 150,668 from 314 subjects.  MEDS-II contains only 836 sample images from 425 subjects imaged in a controlled mugshot setting. Morph-II is a dataset from a longitudinal study that contains 55,063 sample images from 13,000 subjects within a controlled setting over 5 years. While MEDS-II and Morph-II datasets include uncontrolled illumination, Multipie includes controlled illumination samples. ITA, RSR, and SREDS scores were generated for all samples of each dataset.

\begin{figure}[ht]
    \centering
    \includegraphics[width=8cm]{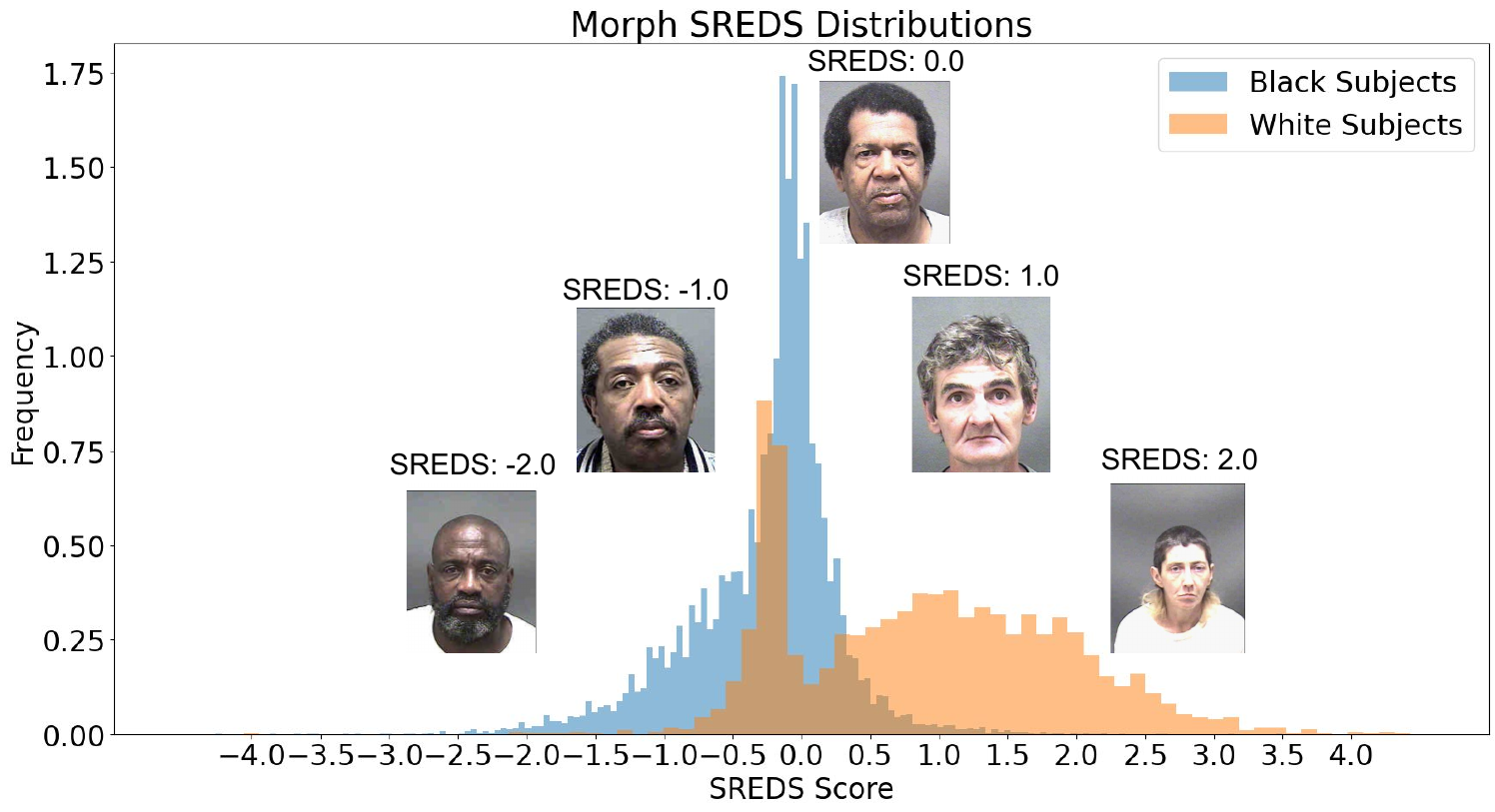}
    \caption{Distribution of SREDS Scores for Morph-II dataset separated by Race. Subjects with specific SREDS scores are shown for the range across the distribution.}
    \label{Morph dist}
\end{figure}

\subsection{Cross-Dataset Analysis}
In prior work \cite{sreds}, the intra-subject variance was used as a metric to describe the variance of a specific subject's skin tone score across multiple samples. The low intra-subject variance shows the metric can produce a consistent value of the same subject independent of external conditions. An example of intra-subject variation can be seen in figure \ref{Intrasubject Variation}. We evaluate and compare the intra-subject variance across all of our evaluation datasets and compare it to other methods. In addition, we test the generalizability of the SREDS metric by calibrating the skin tone gradient on one dataset and applying it to another, thereby testing its robustness to unseen datasets.
The same experiment is run using the RSR PCA models for comparison. ITA does not have a training component and is only reported per dataset.

\subsection{SREDS Agnostic Facial Recognition Model}
To show the potential of SREDS for use in the replacement of race labels, we compared the outcome performance of SREDS versus ground-truth race labels when incorporated into SensitiveNets, a sensitive information removal network \cite{morales2020sensitivenets}.
SensitiveNets provides a novel privacy-preserving neural network feature representation to suppress the sensitive information of a learned space while maintaining the utility of the data. We reimplemented the sensitive removal network as our model for analysis of the suppression of race and skin tone. A diagram of our model is seen in figure \ref{snetsModelImpl}. SensitiveNets contains sensitive information removal dense layers added on top of a pre-trained face recognition backbone. Within our testing, we used a Resnet50 model pre-trained on VGGFace2 as the backbone, consistent with the cited literature \cite{morales2020sensitivenets,vggface2, Xie}. The model's loss function requires a race classifier that acts as the sensitive information detector. The softmax probability from this detector describes the amount of racial information present within a subject's template and the goal of the loss function is to remove the sensitive race information and trend the classifier towards 50\% accuracy. In our experiments, this classifier is either trained on race labels or SREDS scores binned into predetermined groups. The sensitive information removal $\varphi$ layers are then added and trained sequentially using an adversarial approach of triplet loss and an adversarial sensitivity regularizer loss which reduces the amount of sensitive race information from the embedding space. An in-depth look at the model and loss function can be found in the SensitiveNets literature \cite{morales2020sensitivenets}.

\section{Experiment Results}

Our experiments were performed to analyze how the consistency of SREDS performed relative to other skin tone metrics and the outcome of replacing race labels with SREDS-generated labels in a privacy preservation method.

\subsection{Cross-Dataset Analysis Results} 
We performed the cross-dataset analysis of the two skin tone metrics described in Section 2 and SREDS across the three datasets listed in Section 3. We generated ITA, RSR, and SREDS for all subjects from the mentioned datasets. As part of background normalization, RSR assumed consistent lighting, the same acquisition camera, and a constant background. Only the Multi-PIE dataset meets all conditions. However, due to the lack of constant background in MEDS-II and MORPH-II, the background normalization step was bypassed for these datasets. ITA is a non-trainable method so we collected the ITA values from each subject of each dataset. To test SREDS consistency on unseen data we used the Kernal Principal Component Analysis (KPCA) fit to one dataset's diffuse components and used it to transform another dataset's diffuse components. The same process was recreated using the RSR PCA models on the same datasets' selected skin pixel values in order to compare these two methods. 

\begin{table}[ht]
\setlength{\tabcolsep}{2.5pt}
\centering
\begin{tabular}{c|ccccccccc|}
\cline{2-10}
 &
  \multicolumn{9}{c|}{Testing Dataset} \\ \cline{2-10} 
 &
  \multicolumn{3}{c|}{Morph-II} &
  \multicolumn{3}{c|}{MEDS-II} &
  \multicolumn{3}{c|}{Multi-Pie (Mugshot)} \\ \hline
\multicolumn{1}{|c|}{Training Dataset} &
  \multicolumn{1}{c|}{SREDS} &
  \multicolumn{1}{c|}{RSR} &
  \multicolumn{1}{c|}{ITA} &
  \multicolumn{1}{c|}{SREDS} &
  \multicolumn{1}{c|}{RSR} &
  \multicolumn{1}{c|}{ITA} &
  \multicolumn{1}{c|}{SREDS} &
  \multicolumn{1}{c|}{RSR} &
  ITA \\ \hline
\multicolumn{1}{|c|}{Morph-II} &
  \multicolumn{1}{c|}{0.419} &
  \multicolumn{1}{c|}{0.539} &
  \multicolumn{1}{c|}{0.645} &
  \multicolumn{1}{c|}{0.681} &
  \multicolumn{1}{c|}{0.493} &
  \multicolumn{1}{c|}{N/A} &
  \multicolumn{1}{c|}{0.157} &
  \multicolumn{1}{c|}{0.468} &
  N/A \\ \hline
\multicolumn{1}{|c|}{MEDS-II} &
  \multicolumn{1}{c|}{0.457} &
  \multicolumn{1}{c|}{0.540} &
  \multicolumn{1}{c|}{N/A} &
  \multicolumn{1}{c|}{0.463} &
  \multicolumn{1}{c|}{0.493} &
  \multicolumn{1}{c|}{\textbf{0.448}} &
  \multicolumn{1}{c|}{0.186} &
  \multicolumn{1}{c|}{0.470} &
  N/A \\ \hline
\multicolumn{1}{|c|}{Multi-Pie (Mugshot)} &
  \multicolumn{1}{c|}{\textbf{0.399}} &
  \multicolumn{1}{c|}{0.538} &
  \multicolumn{1}{c|}{N/A} &
  \multicolumn{1}{c|}{0.674} &
  \multicolumn{1}{c|}{0.493} &
  \multicolumn{1}{c|}{N/A} &
  \multicolumn{1}{c|}{\textbf{0.138}} &
  \multicolumn{1}{c|}{0.304} &
  0.401 \\ \hline
\end{tabular}
\caption{Cross dataset intra-subject variability analysis between SREDS, RSR, and ITA skin tone metrics. Bolded values are the lowest recorded intra-subject variability in that testing dataset. SREDS scores result in the least variable metric from Morph-II and MultiPie datasets and the second least variable metric in MEDS-II, behind ITA.}
\label{crossdatasettable}
\end{table}

We computed the intra-subject variability of each dataset's skin tone metrics by calculating the standard deviation of each subject's individual skin tone measures and averaging across the dataset. The results of this analysis are seen in Table \ref{crossdatasettable} and suggest that the learning-based algorithms (RSR and SREDS) perform better than ITA when evaluated on the dataset they are calibrated on. 
Viewing our cross-dataset results, we observe that in larger datasets (Morph, Multi-pie), SREDS outperforms both ITA and RSR even when calibrated on a different dataset, suggesting the generalizability of this approach.

\begin{figure}[ht]
    \centering
    \includegraphics[width=8cm]{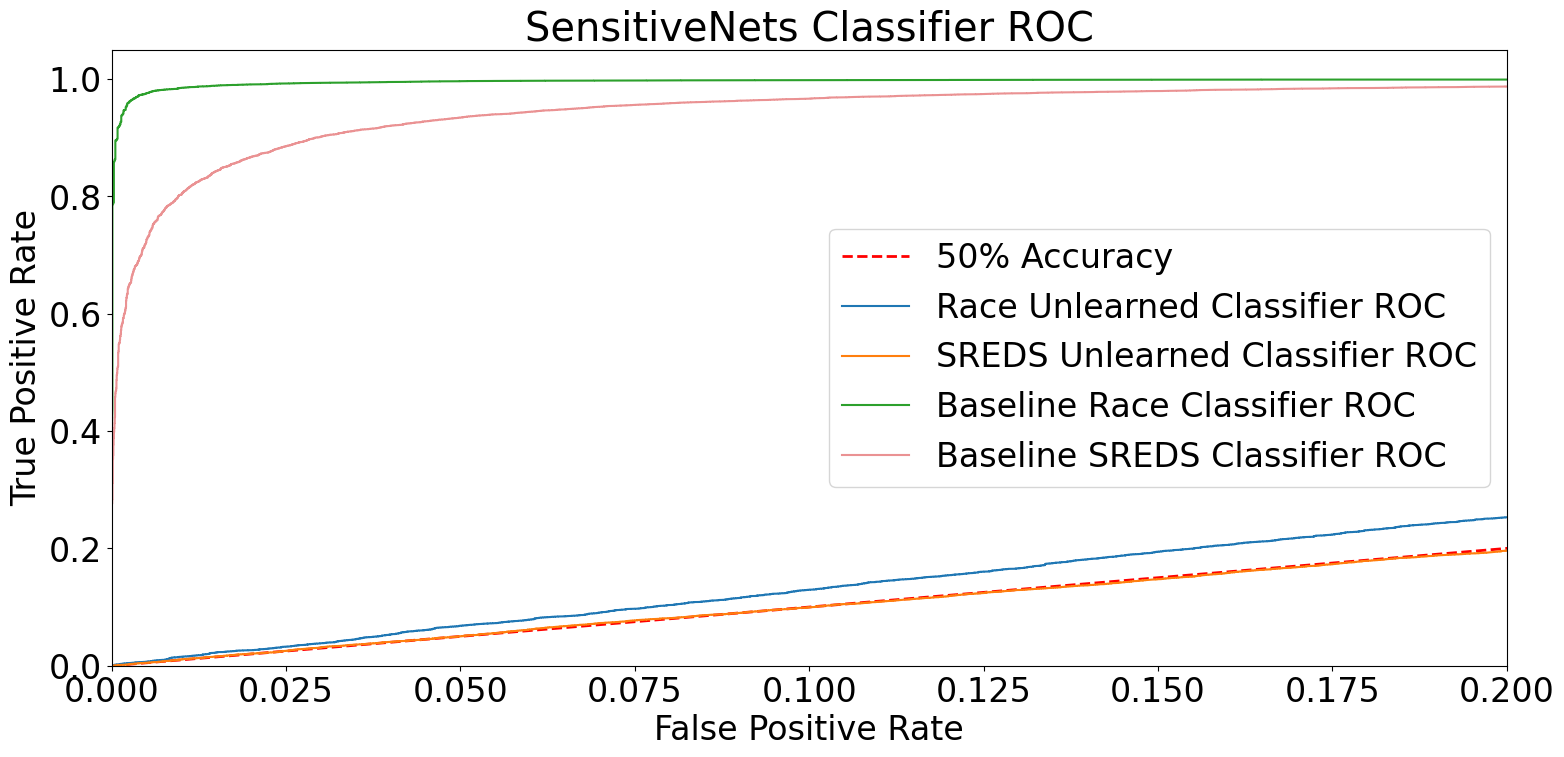}
    \caption{SensitiveNets Sensitive Information Classifier ROCs for both race labels and SREDS scores before and after training on Morph-II dataset. The goal of SensitiveNets training is for 50\% classification accuracy. The unlearned classifier accuracy for both the RDM and SDM is nearly 50\%, which shows SREDS scores and race labels perform similarly in this experiment.}
    \label{snetsClassifier}
\end{figure}

\subsection{Distribution of SREDS}
To utilize SREDS by replacing race labels we needed a process to convert continuous SREDS scores into discrete labels. To understand the distribution of scores, the SREDS scores across the Morph-II dataset were plotted within Figure \ref{Morph dist}.
We split the dataset in half by the median SREDS score of -0.01 and binned the subjects into low and high SREDS scores to create a discrete labeling of the Morph-II dataset.

\subsection{Comparison of Race Labels and SREDS in Sensitive Feature Unlearning}
To see the effects of SREDS scores being used in place of self-reported race labels, we implemented two SensitiveNets models. One model is trained using the black and white subject race labels from the Morph-II dataset while the second model is trained using the binned SREDS value for the same subjects.

\begin{table*}[ht]
\begin{center}
\begin{tabular}{|l|c|c|c|c|c|c}
\hline
Backbone & Classifier & Trained On & Tested On & ICA  & FCA\\
\hline\hline
Resnet50 & Race  & Race Triplets & Morph & 0.985 & 0.47 \\
\hline
Resnet50 & SREDS  & SREDS Triplets & Morph & 0.937 & 0.48 \\
\hline
\end{tabular}
\end{center}
\captionsetup{justification=centering}
\caption{Sensitive Information Removal Network Experiment Results \\ ICA: Initial Classification Accuracy, FCA: Final Classification Accuracy (Goal of sensitive information removal is for FCA to be 0.50)}
\label{Snets table}
\end{table*}

The first model trained on race labels and the second model trained on SREDS scores will be referred to as the Race Unlearned Model (RUM) and the SREDS Unlearned Model (SUM) respectively. An outline of this testing plan is seen in Table \ref{Snets table} with the initial and final classification accuracy of the SensitiveNets classifiers. For both models, the sensitive information classifier ROCs were calculated and shown in Figure \ref{snetsClassifier}. 

\begin{figure}[h!]
    \centering
    \subfloat[ROC comparison for baseline (Resnet50), RUM, and SUM models]
    {\includegraphics[width=3.85cm]{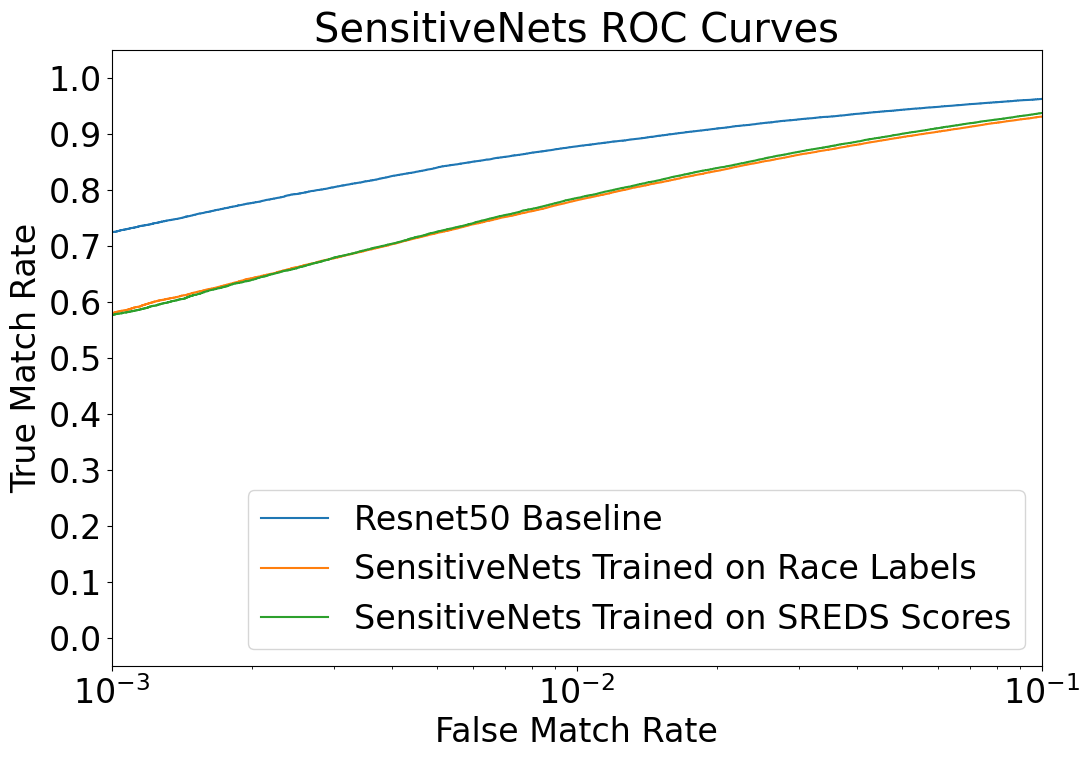}\label{snetsroccomp}}\hspace{.25cm}
    \subfloat[ROC comparison between baseline and RUM model for White and Black subjects.]{\includegraphics[width=3.85cm]{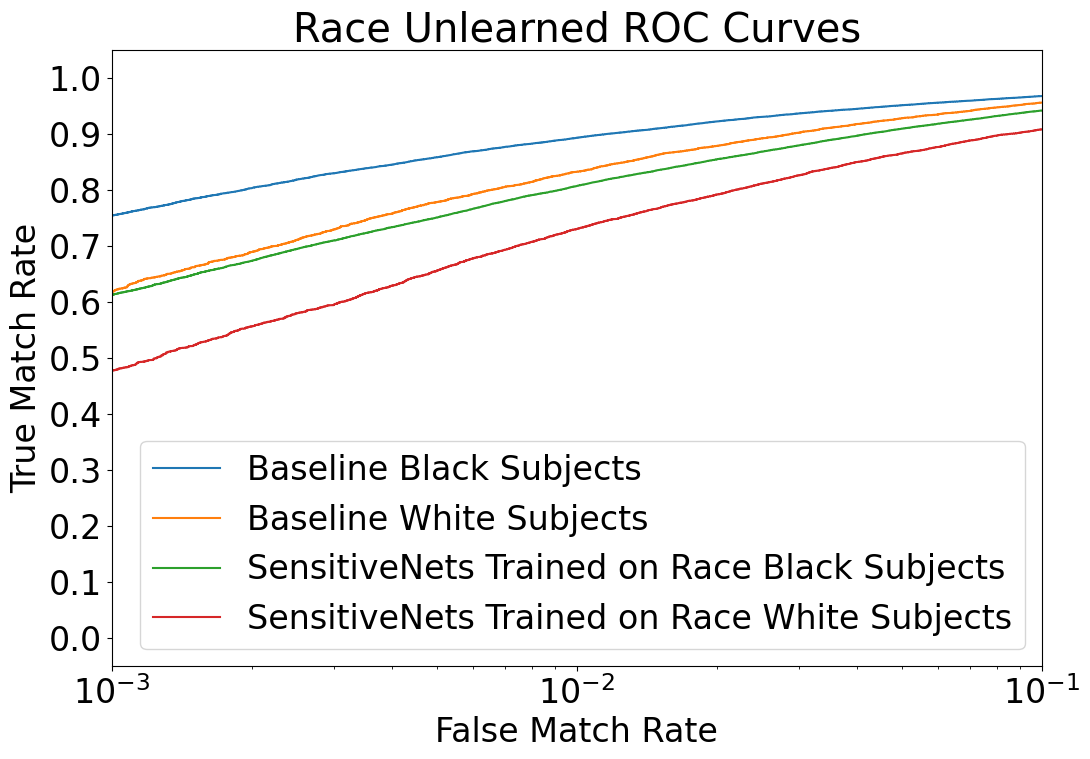}\label{snetsracerroccomp}}\hspace{.25cm}
    \subfloat[ROC comparison between baseline and SUM model for White and Black subjects.]{\includegraphics[width=3.85cm]{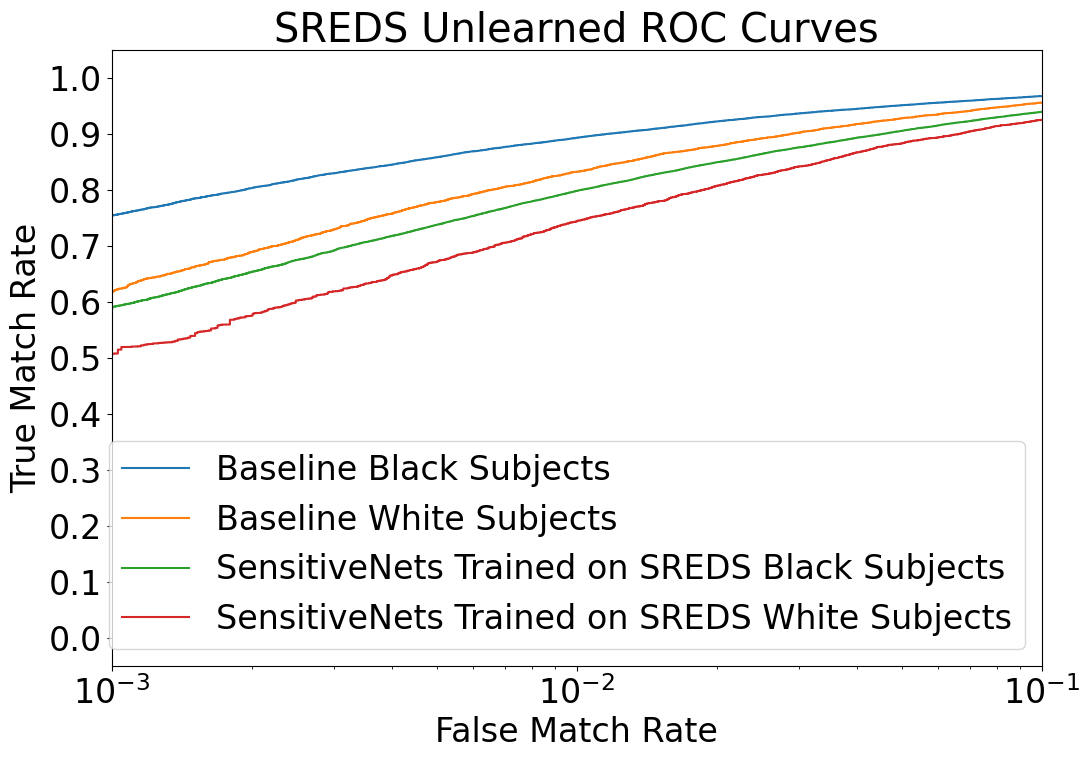}\label{snetssredsrroccomp}}
    \caption{Comparison of biometric performance (matching) ROCs of baseline, RDM, and SDM, categorized by race labels on Morph-II dataset. Shows RDM and SDM suffer from a similar drop in matching performance when race or skin tone information is removed, respectively. }
    \label{snets rocs}
\end{figure}

The two trained SensitiveNets models matching performances are compared to the baseline Resnet50 matching performance to evaluate the results of the training on matching performance in Figure \ref{snets rocs}.
The feature unlearning experiments to preserve privacy show a similar drop in performance between training with race labels and training with SREDS scores. The results suggest only a small (0.027) difference between the True Positive Rate (TPR) of RUM and SUM at $10^{-3}$ False Positive Rate (FPR).  %

\section{Conclusions}
The existing feature unlearning methods in FR rely on large-scale and expensive-to-collect demographically-labeled datasets. Within this study, we demonstrate the ability of SREDS to mitigate this reliance by automatically extracting consistent skin tone information from face images. We have shown that SREDS outperforms other available
skin tone metrics in producing continuous and less-variable skin tone estimates while generalizing well to unseen data. We have presented an application of extracted SREDS scores in the absence of race labels in a feature unlearning method and shown that SREDS could be used as a replacement.

\subsection{Limitations and Future Work}
Limitations of this work include our analysis of only black and white subjects due to the under-representation of other races in our datasets. This led to us only using two SREDS bins when categorizing our datasets to match the binary race labels. We tested using only one face matcher within our privacy-preserving method and have not seen how different networks affect our results. A limitation of using skin tone as a way to label datasets is that skin tone does not encapsulate the entirety of a self-reported race label. Skin tone is one physical characteristic that makes up race and cannot be used as an exact replacement.

Future work planned includes further analysis of the mapping of SREDS to multi-race demographic information and its use in different downstream biometric tasks, recreating our experiments with addition face matches \cite{Deng_2022}, and attempting a bias mitigation solution using SREDS scores and evaluating using fairness metrics \cite{FDR, GARBE} on an even larger scale dataset (BUPT-Globalface) \cite{BUPT}.

\bibliography{lniguide}

\end{document}